\documentclass[fleqn,twoside]{article}
\usepackage{espcrc2}

\usepackage{graphicx}
\usepackage[figuresright]{rotating}

\newcommand{\AmS}{{\protect\the\textfont2
  A\kern-.1667em\lower.5ex\hbox{M}\kern-.125emS}}

\title{Supernova neutrinos, from back of the envelope to supercomputer}

\author{Christian Y. Cardall\address{Physics Division, Oak Ridge National Laboratory, \\
      Oak Ridge, TN 37831-6354, United States of America}%
      \address{Department of Physics and Astronomy, University of Tennessee, \\
	Knoxville, TN 37996-1200, United States of America}%
        \thanks{Oak Ridge National Laboratory is managed by UT-Battelle, LLC, for the DoE under contract DE-AC05-00OR22725.}}

\begin{document}

\begin{abstract}
While an understanding of supernova explosions will require sophisticated large-scale simulations, it is nevertheless possible to outline the most basic features of the neutrino emission resulting from stellar core collapse with a pedestrian account that, through reliance upon broadly accessible physical ideas, remains simple and largely self-contained. 
\vspace{1pc}
\end{abstract}

\maketitle

\section{BACK OF THE ENVELOPE}

Because the central features of a system can often be elucidated with simple models, and because neutrino emission is---from nature's point of view of raw energetics---the central feature of core-collapse supernovae, one may hope 
to understand the basic features of supernova neutrino emission quickly and easily.\footnote{A self-contained but simple derivation will be carried on in these footnotes. In terms of nuclear physics, weak interactions, and hydrodynamics relevant to stellar collapse and bounce, as well as spatial neutrino diffusion during post-bounce deleptonization, this presentation exhibits a cavalier disregard for detail that borders on criminal negligence. For responsible treatments of these aspects that nevertheless retain a somewhat analytic flavor (though relying to an extent on numerical simulations or experimental information), see for instance Refs. \cite{bethe79,fuller82,fuller85,yahil83,brown82,bethe80,bethe82,burrows81}.}

Shortly after the discovery of the neutron in the early 1930s Baade and Zwicky declared: 
``With all reserve we advance the view that supernovae represent the
transitions from ordinary stars to {\em neutron stars,} which in their final
stages consist of extremely closely packed neutrons'' \cite{baade34}. 
For the `core-collapse' varieties of supernovae (Types Ib/Ic/II) this basic picture prevails today with strong theoretical and observational support. 

The story\footnote{In slashing through this problem, the virial theorem \cite{collins78} will prove to be a particularly sharp machete:
\begin{equation}
2E_{N} + E_{R} = -E_G, \label{virial}
\end{equation}
where $E_{N}$ and $E_{R}$ are the contributions to a star's internal energy in nonrelativistic and ultrarelativistic particles respectively, and 
\begin{equation}
E_{G} = -\frac{3}{5}\frac{G M^2}{R} \label{gravity}
\end{equation}
is the total gravitational energy of a star of uniform density (assumed here throughout) with mass $M$ and radius $R$. (In this presentation $G$ is Newton's constant; moreover, $\hbar=c=k=1$.)} 
of how an `ordinary star transitions to a neutron star,' to paraphrase Baade and Zwicky, involves the emission of another electrically neutral particle `discovered' (theoretically, if not experimentally) in the 1930s: the neutrino. Once conditions for their production are reached, the weakness of neutrino interactions implies that their emission is the most efficient means of cooling. Neutrino emission is responsible for the degeneracy of the  precollapse stellar core of mass $M\approx 1.5\:M_\odot$ (the Chandrasekhar mass); the core's eventual instability, at least in part; and the energy loss of about $1.7\times10^{53}\:{\rm erg}$ that ultimately allows collapse to proceed to radius $R_{\rm cold}\approx 11\:{\rm km}$ as a cold neutron star.\footnote{The assumption that the stellar core burns all the way to $^{56}_{26}$Fe---the top of the binding energy curve, from which no further energy can be extracted by fusion---allows an iron core temperature $T_{\rm Fe}\approx 0.9$ MeV to be estimated as follows. Fusion requires tunneling through a Coulomb barrier. Take the uncertainty principle $\Delta x\, \Delta p \approx 1$ as a measure of the potency of tunneling, with $\Delta p \approx \sqrt{2\mu Z_1 Z_2 \alpha / \Delta x}$ as a momentum scale corresponding to the energy height of the Coulomb barrier (here $\mu$ is the reduced mass of colliding nuclei of proton numbers $Z_1$ and $Z_2$, and $\alpha\approx 1/137$ is the fine structure constant). Take also $T\approx Z_1 Z_2 \alpha / \Delta x$ as the characteristic thermal energy per particle needed to bring the colliding nuclei within $\Delta x$ of each other. Elimination of $\Delta x$ yields $T= 2 Z_1^2 Z_2^2 \alpha^2 \mu$. Burning tends to proceed by successive capture of $\alpha$ particles (which have modest $Z$, and are tightly bound and therefore abundant), so taking $Z_1 = 24$, $Z_2=2$, and $\mu = m_N (4)(52)/(4+52)$ appropriate to a putative final $\alpha$ capture to $^{56}_{26}$Fe gives $T_{\rm Fe}\approx 0.9$ MeV ($m_N\approx 939$ MeV is a generic nucleon mass).

If the electrons are not already degenerate at this point, they must soon become so, thanks to neutrino emission. A temperature $T_{\rm Fe}\approx 0.9$ MeV is almost twice the electron mass $m_e$, so in the absence of an electron chemical potential of similar magnitude positrons will be thermally produced. But these $e^+$ will annihilate with $e^-$ into $\nu\bar\nu$ pairs that escape freely; hence we expect cooling by $\nu\bar\nu$ emission back down towards $T\approx m_e \approx 0.5$ MeV. In response to the resulting loss of pressure support the core contracts to higher density, with the loss of gravitational potential energy increasing the temperature and bringing about further pair emission, and so on. Continued increase in density with temperature regulated to $\sim 0.5-1$ MeV eventually results in electron degeneracy, which becomes the dominant source of pressure support. (Iron core central temperatures of $\sim 0.5-1$ MeV are in fact seen in stellar evolution codes \cite{woosley02}.)

The consequences of degeneracy are dire. As the mass and density of the iron core increase during the final burning stage, the velocity of the degenerate electrons approaches the speed of light, at which point they can provide no further pressure support. To estimate the core's maximum mass, take $E_N = 0$ and $E_R = (E_F)_e$ in Eq. (\ref{virial}), where
\begin{equation}
(E_F)_e = \frac{3\pi^{1/3}}{4}\left(3\over 2\right)^{2/3}\left(M\:Y_e\over m_N\right)^{4/3}\frac{1}{R} \label{electrons}
\end{equation}
is the internal energy of a uniform core supported by a degenerate Fermi gas of relativistic electrons; the electron fraction $Y_e$ is the ratio of the net number density of electrons $n_{e^-}-n_{e^+}$ (or, by charge neutrality, the number density of protons $n_p$) to the total baryon number density $n$. Note the cancellation of radius that then results in Eq. (\ref{virial}), a clear indication that something goes haywire when a star tries to support itself by relativistic degeneracy pressure alone. The resulting expression nevertheless yields an estimate of the mass of this extreme configuration, the so-called Chandrasekhar mass:
\begin{equation}
M \approx 1.5\: M_\odot \left(Y_{e,0}\over 0.46 \right)^2, \label{mass}
\end{equation}
where $Y_{e,0} \approx 0.46$ is the ratio $Z/A$ of the proton and mass numbers of $^{56}_{26}$Fe. This may be taken as an estimate of the mass of the core upon collapse, and of the compact remnant that results therefrom. (Taking inhomogeneous stellar structure into account yields $M \approx 1.22\: M_\odot \left(Y_{e,0}/ 0.46 \right)^2$ \cite{chandrasekhar84}.)

Next consider the role of neutrinos in the ultimate instability of the core. As it approaches the Chandrasekhar mass, one proximate cause of the core's instability---which in fact will be used here to define the onset of collapse---is electron capture, which saps pressure support from the core. Electron capture becomes possible when the electron Fermi energy
\begin{equation}
(\epsilon_F)_e = \pi^{1/3}\left(3\over 2\right)^{2/3}\left(M\:Y_e\over m_N\right)^{1/3}\frac{1}{R} \label{eFermiEnergy}
\end{equation}
rises above the energy threshold required to convert a proton to a neutron.
Crudely considering a nucleus of proton number $Z$ and mass number $A$ as comprising degenerate Fermi gases of nonrelativistic protons and neutrons contained in a volume of radius $r_A = r_N A^{1/3}$ (where $r_N\approx 1.2$ fm), the typical emitted neutrino energy $\bar\epsilon_{\nu_e}$ is not the typical energy of an electron captured from its (ambient) degenerate Fermi sea, because the energy required to lift a bound proton from {\em its} Fermi sea (within the nucleus) to the top of the {\em neutron} Fermi sea (within the nucleus) must first be paid. Hence $\bar\epsilon_{\nu_e} \approx (\epsilon_F)_e - \Delta - [(\epsilon_F)_{n;A,Z} - (\epsilon_F)_{p;A,Z}]$ (where $(\epsilon_F)_{n;A,Z}$ and $(\epsilon_F)_{n;A,Z}$ are the Fermi energies of the neutrons and protons bound in the nucleus), or
\begin{eqnarray}
\bar\epsilon_{\nu_e} \!\!\!&\!\!\!\approx \!\!\!&\!\! (\epsilon_F)_e\! -\!\Delta\! -\! \frac{3}{4}\!\left(3\pi^2\over 2\right)^{1/3}\!\! \frac{(A-Z)^{2/3}\!-\!Z^{2/3}}{m_N\, r_N^2 \, A^{2/3}} \\
\!\!\!&\!\!\!\approx \!\!\!&\!\! (\epsilon_F)_e\! -\!\Delta\! -\! \frac{3}{4}\!\left(3\pi^2\over 2\right)^{1/3}\!\! \frac{(1-Y_e)^{2/3}\!-\!Y_e^{2/3}}{m_N \, r_N^2}, \label{neutrinoEnergy}
\end{eqnarray}
where $\Delta=m_n-m_p-m_e \approx 0.8$ MeV and it is assumed that all baryons are locked up in nuclei characterized by the same $Z$ and $A$. Electron capture can proceed when the right-hand side is positive; using $(\epsilon_F)_e$ as given in Eq. (\ref{eFermiEnergy}) yields $R_0 \approx 810$ km at the onset of electron capture for $M=1.5\,M_\odot$ and $Y_{e,0} = 0.46$, corresponding to a density of $\rho_0 \approx 1.3\times 10^{9}\ {\rm g/cm}^3$. 
Incidentally, we are now in a position to compute the entropy per bayon of the precollapse core: $s_0 \approx 0.85-1.5$ (corresponding to $T_0 \approx 0.5-1$ MeV) receives contributions from the nuclei of mass number $A=56$, the relativistic degenerate electrons, and photons:
\begin{eqnarray} 
s_{\rm A} \!\!\!&\!\!=\!\!&\!\!\! \frac{1}{A}\!\left\{\!\frac{5}{2}\!+\!\ln\left[\frac{m_A}{\rho_0}\left(m_A T_0\over 2\pi\right)^{3/2}\right]\! \right\}\nonumber\\
&\approx& 0.31-0.33, \label{sNuclei}\\
s_{e^-} \!\!\!&\!\!=\!\!&\!\!\! \pi^2 Y_{e,0} \frac{T_0}{(\epsilon_F)_e} \approx 0.52-1.04, \\
s_{\gamma} \!\!\!&\!\!=\!\!&\!\!\! \frac{4\pi^2}{45}\frac{m_N\:T_0^3}{\rho_0} \approx 0.02-0.14,
\end{eqnarray}
where $m_A \approx m_N A$ is the mass of a nucleus of mass number $A$ and $(\epsilon_F)_e = (3\pi^2\rho_0 Y_{e,0}/m_N)^{1/3}\approx 4.4$ MeV is the initial electron Fermi energy. (As was the case with $T_0$, these values of $\rho_0$ and $s_0$ are in reasonable agreement with the central iron core values produced by stellar evolution codes \cite{woosley02}.) 

Having characterized the precollapse core and determined that it is doomed to collapse---and 
supposing that the degeneracy pressure of nonrelativistic nucleons will halt collapse---what might we expect the final radius $R_{\rm cold}$ of the compact remnant to be, and why do we expect it to consist of neutrons rather than a mixture of neutrons, protons, and electrons? Aware that weak interactions can interchange neutrons and protons, we provisionally take our star of ``extremely closely packed [nucleons]'' to consist of nonrelativistic degenerate Fermi gases of neutrons and protons and a relativistic degenerate Fermi gas of electrons. Hence $E_N = (E_F)_n+(E_F)_p$ in Eq. (\ref{virial}), where
\begin{eqnarray}
(E_F)_n &=& \frac{9}{20}\left(3\pi^2\over 2\right)^{1/3}\frac{[M(1-Y_e)]^{5/3}}{m_N^{8/3}{R}^2}, \label{efn}\\
(E_F)_p &=& \frac{9}{20}\left(3\pi^2\over 2\right)^{1/3}\frac{(M \:Y_e)^{5/3}}{m_N^{8/3}{R}^2}, \label{efp}
\end{eqnarray}
and $E_R = (E_F)_e$ of Eq. (\ref{electrons}). 
A second constraint is provided by chemical (`$\beta$' or `weak') equilibrium, which requires 
\begin{equation}
\mu_n = \mu_p + \mu_e,\label{beta}
\end{equation}
where the chemical potentials $\mu_i$ include particle masses. That is, $\mu_i = m_i + (\epsilon_F)_i$ at zero temperature, with
\begin{eqnarray}
(\epsilon_F)_n &=&\frac{3}{4}\left(3\pi^2\over 2\right)^{1/3} \frac{[M(1-Y_e)]^{2/3}}{m_N^{5/3}{R}^2}, \\
(\epsilon_F)_p &=&\frac{3}{4}\left(3\pi^2\over 2\right)^{1/3} \frac{(M\:Y_e)^{2/3}}{m_N^{5/3}{R}^2}, 
\end{eqnarray}
and $(\epsilon_F)_e$ given by Eq. (\ref{eFermiEnergy}).
Taking $M = 1.5\: M_\odot$, simultaneous solution of Eqs. (\ref{virial}) and (\ref{beta}) as specified above yields $R_{\rm cold} = 11$ km and $Y_{e,{\rm cold}} = 1.0 \times 10^{-2}$. Hence the label `neutron star' is well deserved! The single equation $(E_F)_n  = -E_{G}/2$ yields 
\begin{equation}
R_{\rm cold} = 11 {\rm\; km} \left(1.5\; M_\odot \over M \right)^{1/3} 
\end{equation}
with $Y_e$ neglected. (The resulting baryon number density $n \approx 0.25\; {\rm fm}^{-3}$ is within a factor of two of the value $0.14\; {\rm fm}^{-3}$ derived from from experimental measurements of the size of atomic nuclei, which also feature densely packed nucleons. A neutron star is not unlike an atomic nucleus of astronomical proportions!)

Note that neutrino emission is necessary not only to get collapse started, but also for it to proceed to completion. This is another consequence of the virial theorem: from Eq. (\ref{virial}) we see that the total energy of the precollapse core is $(E_F)_e + E_{G} \approx 0$, while the total energy of the cold neutron star is $(E_F)_n + E_{G} \approx E_{G}/2$.
Hence an amount of energy equal in magnitude to half the final neutron star gravitational energy,
\begin{equation}
E_{\nu} \approx 1.7\times 10^{53}{\rm\; erg}\left(M \over 1.5\; M_\odot\right)^{2}\left(11{\rm\; km}\over R\right),
\end{equation} 
must be lost to neutrinos (and a bit of it to the explosion) before the neutron star can reach its cold final state.}

Basic characterizations of the precollapse core and the cold neutron star to which it collapses may be sufficient to estimate the total energy loss to neutrinos, but additional features of this emission---which flavors are emitted, at what energies, over what time scales---depend crucially on the flow of electron lepton number and the generation of entropy. `Neutronization' via electron capture entails the flow of electron lepton number from degenerate electrons to electron neutrinos, and the thermal energy associated with finite entropy results in the thermal emission of neutrinos and antineutrinos of all flavors.\footnote{In a sense these two themes are complementary, for degeneracy entails low entropy in general, and in particular a net electron number prevents the rise of a positron population that would allow $\nu\bar\nu$ pair emission of all flavors by $e^+e^-$ annihilation.} 


\begin{figure}
\includegraphics[width=3in]{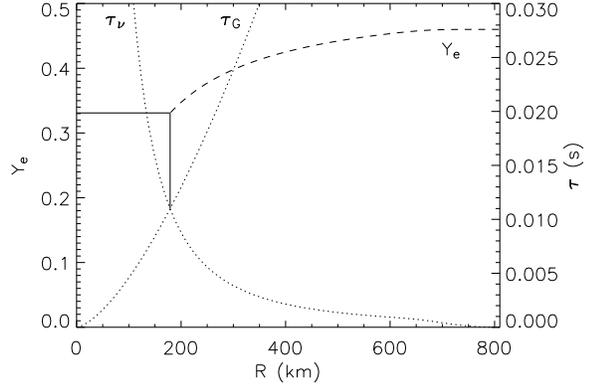}
\vspace{-40pt}
\caption{\small In the crude model of gravitational collapse presented here, the neutrino diffusion time scale $\tau_\nu$ exceeds the free-fall time scale $\tau_G$ when the core radius has shrunk to about 180 km, which corresponds to a trapped lepton fraction of $Y_{e,{\rm trap}} \approx 0.33$.}
\label{leptonFraction}
\vspace{-15pt}
\end{figure}

\begin{figure}
\includegraphics[width=3in]{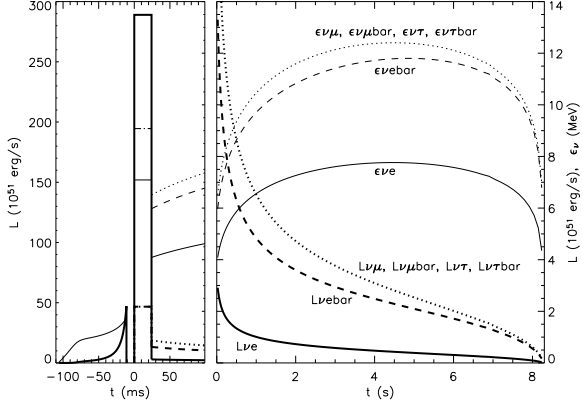}
\vspace{-40pt}
\caption{\small Crudely estimated neutrino luminosities (thick) and characteristic energies (thin). Note the change in time and luminosity (but not energy) scales between the left and right panels. The left panel is a close-up of infall and bounce at $t=0$.}
\label{neutrinos}
\vspace{-15pt}
\end{figure}

The salient point about the flow of electron number during collapse is that while electron neutrinos initially escape freely, they 
eventually become trapped,
as Fig. \ref{leptonFraction} illustrates.\footnote{That neutrino trapping will occur is shown by a comparison of the collapse and neutrino diffusion time scales, but before examining these a word on weak interactions is in order. For energies much lower than the masses of the $W^{+,-}$ and $Z^0$ bosons (whose exchange are responsible for `charged-' and `neutral-current' interactions respectively), the effective weak interaction Lagrangian is
$${\cal L}=-(4G_F/\sqrt{2})(J_{\rm cc})^{\dagger}_\mu (J_{\rm cc})^{\mu} - (G_F/\sqrt{2})(J_{\rm nc})_\mu (J_{\rm nc})^{\mu},$$ where the Fermi constant $G_F = 1.17\times 10^{-11}\ {\rm MeV}^{-2}$ sets the overall scale of weak interaction amplitudes ($\beta$ decay being the prototype), and $(J_{\rm cc})^{\mu}$ and $(J_{\rm nc})^{\mu}$ are a `charged current' (which changes electric charge by one unit) and `neutral current' constructed of quark and lepton fields (e.g. Appendix B of Ref. \cite{kolb90}). In this presentation the factors $8G_F^2$ and $G_F^2/2$ are taken as crude estimates of leading factors in charged and neutral current processes respectively, with neutrino energy factors supplied to give the correct dimensionality for rates or cross sections.

Now to the collapse and neutrino diffusion time scales. Assuming that the collapse attending the core's approach to the Chandrasekhar mass is catastrophic, such that the core is essentially in free-fall, the instantaneous contraction time scale when the core is at density $\rho$ is
\begin{equation}
\tau_G \approx (G \rho)^{-1/2} = \left(4\pi R^3 \over 3 G M \right)^{1/2}. \label{taug}
\end{equation}
On the other hand the trapped electron neutrino diffusion time scale is
\begin{equation}
\tau_{\nu_e} \approx \frac{3R^2}{\lambda_{\nu_e}}\approx \frac{3R^2 \rho\,\sigma_{\nu A\rightarrow A\nu}}{m_N\, A} \approx \frac{9}{8\pi}\frac{A \,G_F^2\, \bar\epsilon_{\nu_e}^2}{m_N}\frac{M}{R}, \label{taunu}
\end{equation}
where $\lambda_{\nu_e}$ is the neutrino mean free path, $\sigma_{\nu A\rightarrow A\nu} \approx A^2\,(G_F^2/2)\,\bar\epsilon_{\nu_e}^2$ is the cross section for coherent neutral current scattering off a heavy nucleus of mass number $A$ (the dominant source of opacity; the argument that the baryons remain locked in heavy nuclei through trapping is given in the next footnote). In accordance with the above remarks, a neutral current scattering cross section per nucleon $(G_F^2/2)\,\bar\epsilon_{\nu_e}^2$ is suggested by dimensional analysis. Applied to an entire nucleus yields not just a factor $A$ corresponding to an incoherent sum over scatterings off individual nucleons, but a factor $A^2$ corresponding to collective scattering off the entire collection of nucleons---a coherent sum in the scattering {\em amplitude}. This is because the neutrino wavelength is larger than the size of the nucleus (as can easily be verified from Eq. (\ref{neutrinoEnergy}) and $r_A = r_N A^{1/3}$ with $r_N \approx 1.2$ fm). With the nucleus then treated as a point particle, the cross section is partially analogous to Rutherford scattering of a charged particle from a point nucleus of proton number $Z$, with our factor of $A^2$ associated with `weak charge number' $A$ corresponding to the factor $Z^2$ in the Rutherford scattering formula.

As collapse begins, 
\begin{equation}
\tau_G \approx 0.11{\rm\; s} \left( R\over 810{\rm\; km}\right)^{3/2} \left(1.5\; M_\odot \over M \right)^{1/2}, \\
\end{equation}
and $\tau_{\nu_e} \approx 0$ since $\bar\epsilon_{\nu_e} \approx 0$ as given by Eq. (\ref{neutrinoEnergy}) has here been taken to define the onset of collapse. While neutrinos initially escape easily, the decrease in $\tau_G$ and increase in $\tau_{\nu_e}$ as $R$ decreases during collapse imply that at some point neutrinos will be trapped (in addition to the explicit dependence of $\tau_{\nu_e}$ on $R$ in Eq. (\ref{taunu}), see Eq. (\ref{eFermiEnergy}) for the increase in $(\epsilon_F)_e$ and hence $\bar\epsilon_{\nu_e}$ as $R$ decreases).

In order to estimate the trapped lepton fraction, note that as long as the $\nu_e$ from electron capture escape freely, the electron fraction changes according to
\begin{eqnarray}
dY_e &\approx& -Y_e\, \Gamma_{eA\rightarrow A\nu}\; dt \\
&\approx& 8\,Y_e \,G_F^2\, \bar\epsilon_{\nu}^5 \left(3\pi R \over G M \right)^{1/2}\; dR.\label{dYe}
\end{eqnarray}
The factor of $Y_e$ gives the fraction of baryons that are protons; the charged current capture rate per proton on nuclei of mass number $A$, $\Gamma_{eA\rightarrow A\nu} \approx 8\,G_F^2\, \bar\epsilon_{\nu}^5$, has been estimated in accordance with previous remarks (in constrast to neutrino scattering, there is no coherence on the nucleus as a whole); and $dt\approx -d\tau_G$ has been taken. Equation (\ref{dYe}), with Eqs. (\ref{neutrinoEnergy}) and (\ref{eFermiEnergy}), is an ordinary differential equation for $Y_e$ whose solution is shown in Fig. \ref{leptonFraction}. Using this solution in connection with Eqs. (\ref{taug}) and (\ref{taunu}), the trapped lepton fraction is the value of $Y_e$ where $\tau_G = \tau_\nu$. This is illustrated graphically in Fig. \ref{leptonFraction} for $M=1.5\:M_\odot$, $R_0=810{\rm\:km}$, and $A=56$. The result is $Y_{e,{\rm trap}}\approx 0.33$.
} 
A roughly estimated $5.0\times 10^{50}$ erg are released in $\nu_e$ before trapping; the gradual rise in total luminosity and cutoff at trapping, as well as the typical neutrino energy during infall, are depicted in the left panel of Fig. \ref{neutrinos} at negative (i.e. pre-bounce) times.\footnote{The total energy released in $\nu_e$ before trapping is obtained by integrating $dE_{\nu_e}=\bar\epsilon_{\nu_e} (M/m_N) (dY_e/dR) dR$ from the initial radius to the trapping radius. The luminosity during infall is $L_{\nu_e}=dE_{\nu_e}/d\tau_G$, and is plotted in Fig. \ref{neutrinos} as a function of time $t=-\tau_G$. The neutrino energy is given by Eq. (\ref{neutrinoEnergy}). In all of this the solution $Y_e(R)$ as obtained in the previous footnote is used.}

\begin{figure}
\includegraphics[width=3in]{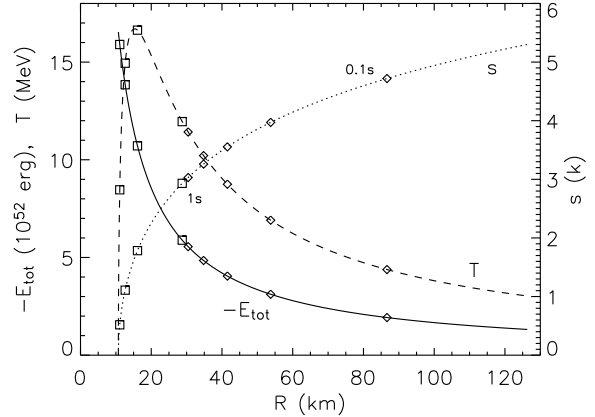}
\vspace{-40pt}
\caption{\small As the nascent neutron star shrinks in radius in response to neutrino emission during the post-bounce cooling phase, the total energy $E_{\rm tot}$ of the core and entropy per baryon $s$ decrease monotonically even as the temperature $T$ increases before making its final plunge. Starting at 0.1 s post-bounce, diamonds mark time intervals of 0.2 s; starting at 1 s, squares mark time intervals of 2 s.}
\label{neutronStar}
\vspace{-15pt}
\end{figure}

The other point to make about collapse is that the entropy per baryon changes only modestly, which conditions what happens on dynamical time scales at the end of implosion.\footnote{To see that the entropy per baryon changes only modestly during collapse, consider the first law of thermodynamics in the form
\begin{equation}
T ds = dq - \sum_i \mu_i dY_i. \label{ds}
\end{equation}
Here $dq = d(e/n) + p\, d(1/n)$ (where $e$ is the internal energy density, $n$ is the baryon number density, and $p$ is the pressure); but it is equal, as the label $dq$ suggests, to the heat per baryon exchanged with the environment. The second term represents the entropy per baryon generated by out-of-equilibrium `chemical' changes to the core ($\mu_i$ are the chemical potentials including rest masses of the species $i$, and $dY_i$ are the changes in the numbers per baryon). There are arguments for the smallness of $ds$ both before and after trapping. 

Before neutrino trapping both energy and lepton number are lost via electron capture, and the modesty of entropy increase results from a substantial cancellation between these contributions. 
The energy per baryon lost to electron capture neutrinos that freely escape is
$
dq\approx  \bar\epsilon_{\nu_e} dY_e. 
$
Meanwhile the chemical change of electron capture involves $dY_e = dY_p = -dY_n$ (the first equality arising from charge conservation and the second from baryon number conservation), so that $\sum_i \mu_i dY_i = (\mu_e + \mu_p - \mu_n)dY_e$. 
Hence
\begin{equation}
T ds \approx (\bar\epsilon_{\nu_e}+ \mu_n - \mu_p  - \mu_e )dY_e. \label{ds2}
\end{equation}
It turns out that the leading factor cancels exactly with the prescription for electron capture neutrino energy of Eq. (\ref{neutrinoEnergy}). In a more careful calculation the cancellation would not be exact, but it would still be substantial. An important consequence of the small change in entropy before trapping is that the baryons remain in nuclei (which feature the coherent scattering cross section $\propto A^2$): 
using $A=1$, $R=180$ km, and $T=1-5$ MeV in Eq. (\ref{sNuclei}) shows that $s\approx 4-6$ would be required for the nuclei to disassemble into free nucleons before trapping. 

It can also be argued that the change in $s$ between trapping and core bounce will be modest, this time because the terms of Eq. (\ref{ds}) tend to vanish individually. First, $dq \approx 0$ since neutrinos can no longer carry away energy. Second, with neutrino trapping, 
\begin{equation}
\sum_i\mu_i dY_i = (\mu_e + \mu_p - \mu_n - \mu_{\nu_e})dY_e + \mu_{\nu_e} dY_L,
\label{chemicalEntropy}
\end{equation} 
where $Y_L = Y_e + Y_{\nu_e}$ is the trapped lepton fraction (in analogy with the definition of $Y_e$ given earlier, $Y_{\nu_e}\equiv (n_{\nu_e} - n_{\bar\nu_e})/n$). 
Between trapping and bounce, while $Y_L = Y_{e,{\rm trap}}$, the second term vanishes because $dY_L \approx 0$, and the first term also becomes small as $e^{-}$, $e^{+}$, $p$, $n$, $\nu_e$ and $\bar\nu_e$ come into chemical (`$\beta$' or `weak') equilibrium. 
}
Instead of a soft landing cushioned by abundant degrees of freedom capable of gently absorbing the kinetic energy of infall, a hard `bounce' can be expected when nucleon degeneracy pressure kicks in at supranuclear densities.\footnote{The estimates of post-bounce neutrino emission presented below involve reasoning from the approximate global states of a neutron star in chemical equilibrium at finite temperature. In addition to thermal photons and $\nu_\mu\bar\nu_\mu$ and $\nu_\tau\bar\nu_\tau$ pairs, there must be allowance for arbitrary degeneracy in $\nu_e\bar\nu_e$ pairs, relativistic $e^+e^-$ pairs, and nonrelativistic neutrons and protons. For relativistic pairs of arbitrary degeneracy there are simple expressions for the differences of number densities and the sums of energy densities of particles and antiparticles, as well as the entropy \cite{bethe80,fuller85}. For the nonrelativistic nucleons the needed quantities can be derived from the Landau potential $\Omega(T,V,\mu) = -p V = E - TS - \mu N$. This can be written in terms of a polylogarithm function as $\Omega(T,V,\mu) = V (m^3 T^5/2\pi^3)^{1/2}{\rm Li}_{5/2}[-\exp(\mu/T)]$, whose symbolic derivatives and numerical evaluation can be handled by any self-respecting symbolic mathematics package (though the Sommerfeld expansion seems more reliable for $\mu/T > 10-20$). The global neutron star states are characterized by two parameters (the entropy per baryon $s$ and the lepton fraction $Y_L=Y_e+Y_{\nu_e}$) and seven unknowns ($R$, $T$, $Y_e$, $\mu_e$, $\mu_{\nu_e}$, $\mu_n$, and $\mu_p$). The seven equations to be solved are the virial theorem of Eq. (\ref{virial}); chemical equilibrium, Eq. (\ref{beta}) with $\mu_{\nu_e}$ added to the left-hand side; the total entropy per baryon; and four `particle number equations' defining $Y_e$, $Y_{\nu_e}=Y_L-Y_e$, $Y_p = Y_e$, and $Y_n = 1-Y_e$ in terms of the chemical potentials, temperature, etc.} 
This is accompanied by a burst of $\nu_e$ emission (see the left panel of Fig. \ref{neutrinos}), involving an energy loss of about $5.9\times 10^{51}$ erg and the sudden release of trapped electron lepton number.\footnote{A neutron star state (as described in the previous footnote) with $s=1.5$ and $Y_L=Y_{e,{\rm trap}}$ has a total energy $E_{\rm tot}$ much lower than $-E_{\nu_e,{\rm infall}}$ (the energy lost to $\nu_e$ escape during infall), so it cannot be the post-bounce state. (As noted earlier, the precollapse initial condition supported by purely relativistic energy is characterized by $E_{\rm tot}=0$, which follows from Eq. (\ref{virial}).) Examining a sequence of states with higher $s$ in a bid to reach $E_{\rm tot}=-E_{\nu_e,{\rm infall}}$ eventually yields a `burst state' with $s\approx 5.9$ whose radius $R\approx 303$ km is equal to its electron neutrino trapping radius. (As before, the neutrino trapping radii are given by equating a diffusion time scale $\tau_\nu = 3 R^2/\lambda$ with a dynamical time scale $\tau_{\rm dyn} = (G\rho)^{-1/2}$. Here $\lambda_{\nu_e}\approx(n_N \sigma_N+ n_n\sigma_n)^{-1}$, where $\sigma_N\approx G_F \bar\epsilon_{\nu_e}^2/2$ is for neutral current scattering on nucleons and $\sigma_n\approx 8G_F \bar\epsilon_{\nu_e}^2$ is for charged current absorption on neutrons. For later use, note that $\lambda_{\bar\nu_e}\approx(n_N \sigma_N+ n_p\sigma_p)^{-1}$, where $\sigma_p\approx \sigma_n$, and $\lambda_{\nu_\mu,\bar\nu_\mu,\nu_\tau,\bar\nu_\tau}\approx(n_N \sigma_N)^{-1}$.) Hence as the core rebounds to this radius, $\nu_e$ with $\bar\epsilon_{\nu_e}\approx 3\mu_e/4 \approx 7.1$ MeV escape freely in a burst in which a lepton fraction $dY_L$ and energy $E_{\nu_e,{\rm burst}}\approx \bar\epsilon_{\nu_e} (M/m_N)dY_L$ are lost during a time $\tau_{\rm dyn}\approx 24$ ms that characterizes this `burst state.' In order to find $dY_L$, a sequence of states is searched by decreasing $Y_L$ while adjusting $s$ to stay just below the (varying) $\nu_e$ trapping radius. All such states are found to be characterized by $E_{\rm tot} < -E_{\nu_e,{\rm infall}} - E_{\nu_e,{\rm burst}}$, all the way down to $Y_{\nu_e}\approx 0$. The conclusion is that in this crude model the deleptonization accompanying the $\nu_e$ burst is essentially complete, with $E_{\nu_e,{\rm burst}} \approx \bar\epsilon_{\nu_e} (M/m_N)Y_{e,{\rm trap}}$ evaluated at the initial `burst state.'   
} 
Moreover, an additional $6.8\times 10^{51}$ erg or so is lost from the core before the dust settles after a dynamical time scale of about 24 ms post-bounce;\footnote{At the end of the sequence of states with declining lepton number searched as described in the preceding footnote, we still do not have a state with $E_{\rm tot}=-E_{\nu_e,{\rm infall}} - E_{\nu_e,{\rm burst}}$; further energy loss is needed. But it is found at the end of this searched sequence that the population of thermal neutrinos of all flavors has begun to make a nontrivial contribution to $E_{\rm tot}$, so their emission is now a viable means of prompt energy loss so long as the star is outside the trapping radius of these flavors. Hence our `post-bounce state' after the dust settles is that characterized by $Y_L=Y_e$ and the highest entropy per baryon that keeps the star within the smallest trapping radius (that of $\nu_{\mu,\tau}$ and their antiparticles, which as described in the previous footnote have the largest mean free path). This is the state with $s\approx 5.3$ and $R\approx 130$ km at the far right of Fig. \ref{neutronStar}. The additional energy loss referred to in the main text is the difference between $E_{\rm tot}$ of this state and $-E_{\nu_e,{\rm infall}} - E_{\nu_e,{\rm burst}}$. 
} 
in the left panel of Fig. \ref{neutrinos} this full amount is attributed to neutrino emission equally divided among all flavors.\footnote{Some fraction of this would actually be transferred on a dynamical time scale to layers exterior to the core---in an incipient `prompt' explosion, it was (rashly) hoped in bygone days---by processes that cannot be discussed in detail here, such as the direct momentum impulse of the rebounding core, `$p\,dV$' work, shock heating, and so on.} 

The neutrino emission encountered thus far accounts for only about 10\% of the total energy to ultimately be lost, with the rest---about $1.5\times 10^{53}$ erg---temporarily stored as heat, thanks to the entropy generated by bounce and its aftermath. As the hot nascent neutron star shrinks (see Fig. \ref{neutronStar}), this energy is lost to neutrino emission of all flavors (albeit with a modest hierarchy of luminosities and average energies) over a period of many seconds, as depicted in Fig. \ref{neutrinos}---a recognizable if very rough caricature of light curves produced by detailed models \cite{buras06,liebendoerfer04,thompson03,prakash01a}.\footnote{The sequence of neutron star states in Fig. \ref{neutronStar} is that characterized by $Y_L=Y_e$ and $s$ stepped down from 5.3 to 0. The luminosities $L_{\nu_i}$ and average energies $\bar\epsilon_{\nu_i}$ for the various species are found by equating two different expressions for the neutrino luminosities necessary to allow a transition between two successive neutron star states. Defining $dE_{\rm tot}$ to be the difference in $E_{\rm tot}$ between two successive states in the sequence, and assuming that the star contrives to release an equal amount of energy in each flavor, one expression is the ratio of $-dE_{\rm tot}/6$ to the neutrino diffusion time scale $\tau_\nu$ (given for the various species in a previous footnote) at that point in the sequence. A second expression for luminosity is $L_{\nu_i}=4\pi R^2(7\pi^2/960)(\bar\epsilon_{\nu_i}/3)^4$, a neutrino version of the Stefan-Boltzmann law with $T_{{\rm eff},i}\approx \bar\epsilon_{\nu_i}/3$. Once the $\bar\epsilon_{\nu_i}$ and $L_{\nu_i}$ are determined, the sequence is related to a time coordinate by defining $dt\equiv -dE_{\rm tot}/(\sum_i L_{\nu_i})$ for each step of the sequence.
}

\section{SUPERCOMPUTER}
\label{sec:survey}

That some basic features of supernova neutrino emission can be motivated by a stick-figure-quality model provides a baseline degree of instant gratification. This is in contrast to the explosion manifest in an expanding supernova remnant's kinetic energy, which, as a 1\% subsidiary detail from nature's perspective, has been requiring decades of detailed study by numerous people to understand \cite{woosley05}.\footnote{Such is the (probably unavoidable) result of our anthropocentric chauvinism. For decades, and not yet sure even now the extent to which the end is in sight, we have obsessed over the explosion mechanism to a greater extent than the neutrino emission, with our detailed interest in the latter arising primarily to the extent it seems useful for the former. There are at least two self-centered reasons for this fixation on the explosion: it ultimately gives rise to the optical display that we can observe with our unaided inborn detection instruments, viz., our eyes; and it is an important source of production and agent of dispersion of the `heavy' elements that, while being so uncharacteristic of the universe's average composition, dominate the composition of our bodies and terrestrial environment.}
 
But stick figures do not ultimately satisfy, of course; we long for the voluptuous contours of a fully-fleshed-out model. In fact, 
under thorough dissection the simple model of the previous section might stand revealed as more retrospectively illustrative than predictively robust.\footnote{This is true also of the more thorough but still somewhat analytic works cited in the first footnote, which to a large extent are in the character of interpreting and illuminating (and in some cases relying upon) features first seen in numerical simulations, rather than being strongly predictive.} This is one indication of the ultimate necessity of detailed simulations. Another is that the details of early neutrino emission and the explosion are more interrelated than some of my previous comments might suggest. 
Whether it be by playing the spoiler, acting as agent of explosion, or setting the conditions necessary to the operation of some other mechanism, neutrinos have played direct or indirect roles in the many explosion scenarios discussed over the years---and detailed simulation of collapse and the second or so after bounce will allow a neutrino signal detected from a Galactic supernova to serve as an unrivaled diagnostic tool. In its full glory neutrino transport is a time-dependent six-dimensional problem, as it requires the tracking of neutrino energy and angle distributions at every point in space. The computational resources necessary to solve the daunting fulness of this problem are still just over the horizon, but efforts to meet this problem in its full measure are already underway \cite{cardall04,hubeny06}.

\def\aap{Astron. Astrophys. }
\def\aasma{Am. Astron. Soc. Meet. Abs. }
\def\apj{Astrophys. J. }
\def\apjl{Astrophys. J. Lett. }
\def\apjs{Astrophys. J. Supp. Ser. }
\def\araa{Annu. Rev. Astron. Astrophys. }
\def\baas{Bull. Am. Astron. Soc. }
\def\jcam{J. Comp. Appl. Math. }
\def\nat{Nature }
\def\natphys{Nature Phys. }
\def\npa{Nucl. Phys. A }
\def\pr{Phys. Rep. }
\def\prd{Phys. Rev. D }
\def\prl{Phys. Rev. Lett. }
\def\prv{Phys. Rev. }
\def\rmp{Rev. Mod. Phys. }


\begin{thebibliography}{10}
\expandafter\ifx\csname url\endcsname\relax
  \def\url#1{\texttt{#1}}\fi
\expandafter\ifx\csname urlprefix\endcsname\relax\def\urlprefix{URL }\fi

\bibitem{bethe79}
H.~A. {Bethe}, G.~E. {Brown}, J.~{Applegate}, J.~M. {Lattimer}, \npa 324 (1979)
  487--533.

\bibitem{fuller82}
G.~M. {Fuller}, W.~A. {Fowler}, M.~J. {Newman}, \apj 252 (1982) 715--740.

\bibitem{fuller85}
G.~M. {Fuller}, W.~A. {Fowler}, M.~J. {Newman}, \apj 293 (1985) 1--16.

\bibitem{yahil83}
A.~{Yahil}, \apj 265 (1983) 1047--1055.

\bibitem{brown82}
G.~E. {Brown}, H.~A. {Bethe}, G.~{Baym}, \npa 375 (1982) 481--532.

\bibitem{bethe80}
H.~A. {Bethe}, J.~H. {Applegate}, G.~E. {Brown}, \apj 241 (1980) 343--354.

\bibitem{bethe82}
H.~A. {Bethe}, A.~{Yahil}, G.~E. {Brown}, \apjl 262 (1982) L7--L10.

\bibitem{burrows81}
A.~{Burrows}, T.~J. {Mazurek}, J.~M. {Lattimer}, \apj 251 (1981) 325--336.

\bibitem{baade34}
W.~{Baade}, F.~{Zwicky}, \prv 45 (1934) 138.

\bibitem{collins78}
G.~W. {Collins}, II, {The virial theorem in stellar astrophysics}, Tucson,
  Ariz., Pachart Publishing House (Astronomy and Astrophysics Series.~Volume
  7), 1978.~143 p., 1978.

\bibitem{woosley02}
S.~E. {Woosley}, A.~{Heger}, T.~A. {Weaver}, \rmp 74 (2002) 1015--1071.

\bibitem{chandrasekhar84}
S.~{Chandrasekhar}, \rmp 56 (1984) 137--147.

\bibitem{kolb90}
E.~W. Kolb, M.~S. Turner, The Early Universe, Vol.~69 of Frontiers in Physics,
  Addison-Wesley, 1990.

\bibitem{buras06}
R.~{Buras}, M.~{Rampp}, H.-T. {Janka}, K.~{Kifonidis}, \aap 447 (2006)
  1049--1092.

\bibitem{liebendoerfer04}
M.~{Liebend{\" o}rfer}, O.~E.~B. {Messer}, A.~{Mezzacappa}, S.~W. {Bruenn},
  C.~Y. {Cardall}, F.-K. {Thielemann}, \apjs 150 (2004) 263--316.

\bibitem{thompson03}
T.~A. {Thompson}, A.~{Burrows}, P.~A. {Pinto}, \apj 592 (2003) 434--456.

\bibitem{prakash01a}
M.~{Prakash}, J.~M. {Lattimer}, J.~A. {Pons}, A.~W. {Steiner}, S.~{Reddy}, in:
  D.~{Blaschke}, N.~K. {Glendenning}, A.~{Sedrakian} (Eds.), LNP Vol. 578:
  Physics of Neutron Star Interiors, 2001, pp. 364--+.

\bibitem{woosley05}
S.~{Woosley}, T.~{Janka}, \natphys 1 (2005) 147--154.

\bibitem{cardall04}
C.~Y. Cardall, in: R.~Rannacher, R.~Wehrse (Eds.), Numerical Methods for 3D
  Radiative Transfer, Springer in press, astro-ph/0404401.

\bibitem{hubeny06}
I.~{Hubeny}, A.~{Burrows}, astro-ph/0609049.

\end{thebibliography}

\end{document}